\documentclass[12pt]{article}
\usepackage[nosort]{cite}
\usepackage{graphicx}
\usepackage[font={scriptsize,singlespacing}]{caption}
\usepackage{epstopdf}
\usepackage{amsmath}
\usepackage{amssymb}
\usepackage{subfigure}
\usepackage{hyperref}
\usepackage{url}
\usepackage{xcolor}
\usepackage{color}
\usepackage{epigraph}

\def\sideremark#1{\ifvmode\leavevmode\fi\vadjust{\vbox to0pt{\vss
 \hbox to 0pt{\hskip\hsize\hskip1em
 \vbox{\hsize2cm\tiny\raggedright\pretolerance10000
 \noindent #1\hfill}\hss}\vbox to8pt{\vfil}\vss}}}%
\definecolor{amaranth}{rgb}{0.9, 0.17, 0.31}
\definecolor{purple(munsell)}{rgb}{0.62, 0.0, 0.77}
\definecolor{americanrose}{rgb}{1.0, 0.01, 0.24}
\definecolor{palatinateblue}{rgb}{0.15, 0.23, 0.89}
\definecolor{royalblue(web)}{rgb}{0.25, 0.41, 0.88}
\definecolor{hanpurple}{rgb}{0.32, 0.09, 0.98}
\definecolor{beaublue}{rgb}{0.74, 0.83, 0.9}
\definecolor{carminered}{rgb}{1.0, 0.0, 0.22}
\definecolor{brightpink}{rgb}{1.0, 0.0, 0.5}
\hypersetup{ linktoc=all,
    colorlinks, linkcolor={palatinateblue},
    citecolor={brightpink}, urlcolor={purple(munsell)}
}

\begin{document}
\thispagestyle{empty}
\begin{center}

\null \vskip-1truecm \vskip2truecm

{\Large{\bf \textsf{The Persistence of the Large Volumes in Black Holes}}}

\vskip1truecm

\textbf{\textsf{Yen Chin Ong}}\\
\vskip0.1truecm
{Nordita, KTH Royal Institute of Technology and Stockholm University, \\ Roslagstullsbacken 23,
SE-106 91 Stockholm, Sweden}\\
{\tt Email: yenchin.ong@nordita.org}\\

\end{center}
\vskip1truecm \centerline{\textsf{ABSTRACT}} \baselineskip=15pt

\medskip

Classically, black holes admit maximal interior volumes that grow asymptotically linearly in time. We show that such volumes remain large when Hawking evaporation is taken into account. Even if a charged black hole approaches the extremal limit during this evolution, its volume continues to grow; although an exactly extremal black hole does not have a ``large interior''. We clarify this point and discuss the implications of our results to the information loss and firewall paradoxes.

\addtocounter{section}{1}
\section* {\large{\textsf{1. The Large Interiors of Black Holes and Information Loss}}}

The concept of ``the volume'' of a black hole is not a well-defined one in general relativity, because 3-volumes depend on the choice of spacelike hypersurfaces [see \cite{NM} for some explicit examples].
Nevertheless, recently Christodoulou and Rovelli \cite{1411.2854} have shown that one can still sensibly talk about the \emph{largest} volume bounded by the event horizon of a black hole. 

For a specific example, they showed that for an asymptotically flat Schwarzschild black hole with ADM mass $M$, the largest volume contained within it grows like
\begin{equation}\label{S}
V \sim 3\sqrt{3}\pi M^2 v
\end{equation}
at late time,
where $v$ is the advanced time defined by
\begin{equation}
v:= t + \int \frac{dr}{f(r)} = t + r + 2M \ln \bigg|\frac{r}{2M}-1\bigg|; ~~f(r) := 1-\frac{2M}{r}.
\end{equation}
This maximal volume corresponds to the $r=3M/2$ hypersurface, which is the maximal spacelike slice in the interior of Schwarzschild geometry \cite{reinhart, 2814}. 
More explicitly,
it is the integral
\begin{equation}\label{volexp}
V \sim \int^v \int_{\Omega} \max_r \left[ r^2 \sin\theta \sqrt{\frac{2M}{r}-1} \right] ~dv d\theta d\phi.
\end{equation}
The lower integration limit of $v$ is omitted since it gives a negligible contribution when we are only concerned with ``large'' volume.
Here, as well as throughout this work, we restrict ourselves to 4-dimensional spacetimes unless otherwise specified. We use the units such that $G=c=k_B=1$, where $G$ and $k_B$ denote the Newton constant and the Boltzmann constant, respectively. We will keep $\hbar$ explicit to emphasize the roles of quantum effects.

The result in Eq.(\ref{S}) means that even though classically the black hole remains static from the exterior point of view --- and so has the same area $16\pi M^2$ forever --- its maximal interior volume grows with time\footnote{See also sec.(6.3.12) in \cite{0308048} in which it was pointed out that ``there is no dearth of space inside [a black hole]''.} [c.f. a different proposal for black hole's volume in \cite{0508108}]. The result was later generalized to the rotating case [asymptotically flat Kerr black holes] by Bengtsson and Jakobsson \cite{1502.01907}. Recently, it was further shown that such a measure of black hole volume does not necessarily behave in accordance to our flat space intuition. Specifically, it need \emph{not} be a monotonically increasing function of the horizon area \cite{oyc}. For example, it was found that for asymptotically locally AdS black holes with toral topology and AdS length scale $L$, the maximal interior volumes grow like
\begin{equation}
V \sim 4 \pi ML v,
\end{equation}
\emph{independent} of the size of their horizons, which depend on an additional parameter $K$ that determines the compactification of the torus\footnote{In 4-dimensions, the black hole horizon is 2-dimensional and one can think of a flat square torus as the product of two circles, $T^2 = S^1 \times S^1$, each of which has period $2\pi K$.}. Furthermore, an asymptotically locally AdS black hole with lens space topology $S^3/\Bbb{Z}_p$ [in 5-dimensional spacetime] continues to have maximal interior volume\footnote{This volume expression is, curiously, also the same as that of 5-dimensional AdS toral black holes \cite{oyc}.}
\begin{equation}
V \sim \frac{8}{3} \pi ML v,
\end{equation}
as we take the limit $p \to \infty$, which shrinks the horizon to zero area.

The idea that black holes have large interiors is certainly not new. For example, it is well known that the maximally extended Schwarzschild [Kruskal-Szekeres] geometry contains an entire universe on the ``other side'' of the event horizon. In addition,
one can simply glue a closed FLRW universe to the ``other side'' of a Schwarzschild black hole via the Einstein-Rosen bridge, a construction known as the Wheeler's ``bag-of-gold'' \cite{Wheeler}. One can even construct ``monster'' --- a \emph{non-black hole} configuration with finite area but arbitrarily large volume in general relativity, although it is unstable \cite{monsters, oyc2}. 

However, what Christodoulou and Rovelli \cite{1411.2854} found is remarkable since their result showed that large volumes exist in any \emph{generic} black hole, even one that was formed from gravitational collapse [and thus lacks a second asymptotic region]. This is especially important in the context of information loss paradox, which is best formulated by considering a pure quantum state that collapses into a black hole, which then Hawking evaporates away. If the Hawking radiation is purely thermal, and if the end state is complete evaporation, then a pure quantum state would have evolved into a mixed state of Hawking radiation, seemingly violating the unitarity of quantum mechanics \cite{hawking}. Indeed, large interiors of various kinds have long been proposed as possible resolutions to the paradox [see \cite{1412.8366} for a recent review]. The basic idea is that the large volume in a black hole should be able to contain all the information, even though the area has shrunk to a very small size. If black hole evaporation eventually stops, i.e. the black hole becomes some kind of ``remnant'' \cite{1412.8366,A}, then there is no information loss since the information safely remains inside the black hole, which has enough room to store said information if it has sufficiently large interior volume. 

It is therefore of paramount importance to investigate the \emph{time evolution} of the maximal volumes contained in black holes, as the holes undergo Hawking evaporation. The point is that, if the large volumes cease to be ``large'' near the end of the evaporation process, then they cannot be responsible for storing information, at least not in the naive way we discussed above. In the following section, we will show that the large interior volumes actually do continue to grow even though the black holes shrink in area.  
We then discuss the charged case --- although the same result holds --- it is important to clarify why this is in principle consistent with the observation that the ``large'' volumes vanish in the extremal limit\footnote{It is worth emphasizing that the lack of maximal ``large volume'' in the sense of Christodoulou and Rovelli, does not imply the absence of \emph{any} volume behind the horizon. Taken at face value, the Penrose diagram for an extremal black hole clearly shows that there is still ``normal'' spacetime region inside [see, however, \cite{1005.2999, 1105.2574}].}.
Finally we discuss the implications of this result to the information loss and firewall paradoxes \cite{amps1, amps2, BPZ}.

\addtocounter{section}{2}
\section* {\large{\textsf{2. The Time Evolution of Neutral Black Hole Volumes}}}

Let us first consider an asymptotically flat Schwarzschild black hole, which gradually evaporates away. 
Note that the Hawking temperature is usually measured with respect to Schwarzschild time $t$. This is evident, for example, in the Wick rotation approach in which one starts from the standard Schwarzschild metric in $(t,r,\theta,\phi)$ coordinates and goes to Euclidean signature to obtain the Hawking temperature by ensuring regularity at the Euclidean horizon [see, e.g., page 412 of \cite{johnson}]. Since the expression of the maximal volume is in terms of the advanced time $v$, it is convenient to assume the Vaidya model \cite{vaidya1} of Hawking evaporation instead [see Section (3.3) of \cite{modeling}], that is $M=M(v)$ in the Eddington-Finkelstein metric:
\begin{equation}
ds^2=-\left[1-\frac{2M(v)}{r}\right]dv^2 + 2dvdr +r^2(d\theta^2 + \sin^2\theta^2 d\phi^2).
\end{equation}
This is an exact solution to the Einstein field equations with stress-energy tensor $T_{vv}=L(v)/(4\pi r^2)$, where $dM/dv=L(v)$ is the radial flux one uses to model Hawking radiation.
[Alternatively, we can work in terms of $t$ by fixing an asymptotic observer at some large $r$, so that $\Delta t =\Delta v$.]

Since Hawking evaporation is thermal\footnote{Excitations over the thermal spectrum are to be expected for many models of Hawking evaporation. However one should not expect a large deviation from a thermal spectrum, at least up to the Page time \cite{page1, page2}. For a pedagogical introduction to the Page time and quantum information theory in the context of black hole information loss, see \cite{harlow}.}, the rate of mass loss is given by the following differential equation\footnote{If black hole evaporation is indeed unitary, it has been argued that the mass loss is \emph{not} monotonic, due to the flux of negative energy that reached asymptotic infinity \cite{1404.0602, 1405.5235}. However such influx of negative energy is strongly constrained by quantum energy inequalities \cite{1404.0602} and therefore should not affect the overall evolution of black holes too much.}:
\begin{equation}\label{ODE1}
\frac{dM}{dv} \approx -\alpha a\sigma T^4 = - \alpha a \left(27\pi M^2\right) \left(\frac{\hbar}{8\pi M}\right)^4 =  -\frac{\mathcal{C} }{M^2},  ~~\mathcal{C} > 0,
\end{equation}
where $a=\pi^2/(15 \hbar^3)$ is the radiation constant [this is $4/c$ times the Stefan-Boltzmann constant], and $\alpha$ is the greybody factor, which depends on the number of particle species emitted by the Hawking radiation.
The precise value of $\alpha$ is not important for our purpose and we have simply denoted by $\mathcal{C}$ the [dimensionful] coefficient of $1/M^2$ in the differential equation. Note that, assuming that we start with a sufficiently large black hole, the Hawking radiation is dominated by massless particles since the Hawking temperature is too low to pair-produce massive ones. This will remain the case for much of the evolution until near the end of the black hole lifetime \cite{page3}. Here, $\sigma$ is the geometric optic cross section of the Schwarzschild geometry [see, e.g., equation (6.3.34) of \cite{wald}], not the area of the horizon [see also, \cite{HW}]. This fact is not too important in this particular case since it is proportional to the horizon area and one may simply absorb the numerical factor into the redefinition of $\mathcal{C}$. However this is not necessarily the case for other black holes, as we shall see below. 

Solving Eq.(\ref{ODE1}), we obtain the mass of the black hole as a function of time:
\begin{equation}\label{Mt}
M(v) \approx \left(M_0^3 - 3\mathcal{C}v\right)^{\frac{1}{3}},
\end{equation}
where $M_0$ is the initial mass of the black hole at $v=0$. [Strictly speaking, we should not take $v=0$ as the time when the black hole was formed, but some time after the volume expression Eq.(\ref{S}) becomes dominant.]
Now, the maximal volume is, from Eq.(\ref{volexp}),
\begin{equation}
V(v) \sim 3\sqrt{3}\pi \int^v M^2 ~dv.
\end{equation}
\emph{Note that the volume is no longer asymptotically linear in $v$, as it would be classically.}

The derivative is then, by the Fundamental Theorem of Calculus,
\begin{equation}
\frac{dV}{dv} \sim  3\sqrt{3} \pi M^2(v) \geqslant 0.  
\end{equation}
Thus the volume increases --- although with ever decreasing rate --- until it reaches its maximal value
at time 
\begin{equation}\label{Smax}
v_{\text{max}} = \frac{M_0^3}{3 \mathcal{C}},
\end{equation}
which is the same as the evaporation time. This means that for an asymptotically flat Schwarzschild black hole, its maximal volume continues to increase even as the horizon area decreases in size. Furthermore, the volume is seemingly finite even when the black hole horizon goes to zero in the limit, at the end stage of Hawking evaporation.

For another concrete example, we now examine the case of an asymptotically locally AdS black hole whose horizon has the topology of a flat torus [for the explicit metric tensor, see, e.g., \cite{oyc,1403.4886,danny}]. 
Due to the boundary of the geometry, large black holes in asymptotically [locally] AdS spacetimes do not evaporate but instead settle down to thermal equilibrium with their Hawking radiation, which get reflected back from infinity in finite time.  Nevertheless it is possible to modify the boundary conditions to allow black holes to evaporate. This can be achieved, for example, by coupling an auxiliary system to the boundary \cite{rocha, mark}. The Hawking temperature of a neutral toral black hole is
\begin{equation}
T=\frac{3\hbar r_h}{4\pi L^2}.
\end{equation} 
For a fixed $L$, there is an upper bound of $M$ such that the black hole is not too hot, and thus only emits massless particles in its Hawking radiation \cite{1403.4886}. 

The effective potential for massless particle in this geometry does not have a local maximum [like in the case of an asymptotically flat Schwarzschild black hole at $r=3M$], but instead tends to a constant $J^2/L^2$, for any fixed angular momentum $J$ of the emitted particle \cite{1403.4886}. Without a local maximum, the notion of ``capture'' is not interchangeable with ``escape''. In fact, every ingoing massless particle reaches the black hole, but not all massless particles can escape. It turns out that the relevant cross section for Hawking emission is proportional to $L^2$, \emph{not} the horizon area \cite{1403.4886,9803061}, and the thermal mass loss differential equation is
\begin{equation}
\frac{dM}{dv} \approx -a\pi^2 K^2 L^2 T^4 = -a\pi^2 K^2 L^2 \left(\frac{3\hbar M}{2\pi^2 K^2 r_h^2}\right)^4 = -\mathcal{B}M^{\frac{4}{3}}, ~~\mathcal{B}>0,
\end{equation}
where
\begin{equation}
r_h = \left(\frac{2ML^2}{\pi K^2}\right)^{\frac{1}{3}}
\end{equation}
is the horizon of the black hole. Again, we only care about the behavior of the mass $M$ and call all other factors $\mathcal{B}$.

The solution is 
\begin{equation}\label{Mt2}
M(v) = \left(\frac{3}{\mathcal{B}v + 3 M_0^{-\frac{1}{3}}}\right)^3.
\end{equation}
Note that unlike asymptotically flat Schwarzschild black holes, which evaporate in finite time, AdS toral black holes only tend to zero mass asymptotically, i.e., as $v \to \infty$. [See also \cite{9803061}.]
The maximal volume of such a black hole is \cite{oyc} 
\begin{equation}
V \sim 4\pi L \int^v M(v) ~dv.
\end{equation}
[The advanced time here of course differs from that of an asymptotically flat Schwarzschild black hole.]

A straightforward calculation shows that its derivative is
\begin{equation}
\frac{dV}{dv} \sim 4\pi L \left(\frac{3}{\mathcal{B}v + 3 M_0^{-\frac{1}{3}}}\right)^3 \geqslant 0
\end{equation}
Again, we see that the volume is monotonically increasing, whereas the horizon size [and the mass] is decreasing asymptotically toward zero.

\addtocounter{section}{3}
\section* {\large{\textsf{3. The Time Evolution of Charged Black Hole Volumes}}}

The examples above showed that the maximal volume inside a neutral black hole remains large even when Hawking evaporation is taken into account. 
This result generalizes to more complicated black holes. To see why, let us consider a generic static black hole. Its maximal volume expression is,  up to some constant factors, 
\begin{equation}
V  \sim \int^v \max_r[f(r, \left\{M_i(v)\right\})]~dv,  
\end{equation}
where $M_i(v)$'s are some parameters of the black holes such as the mass, while $f$ is a positive function. The derivative of $V$ will then always be positive, and so the volume always grows. 
This suggests that there is sufficient room for information storage if the black hole becomes a remnant at the end of the evaporation.

However, 
there remains a subtle point to clarify. It is observed in \cite{1502.01907} that in the case of an asymptotically flat Kerr black hole, the volume [for fixed time] becomes smaller as the ratio $a/M$ increases, and 
vanishes in the extremal limit, i.e. when $a \to M$. This is also true for the charged case. The evolution of a sufficiently large asymptotically flat Reissner-Nordstr\"om black hole under Hawking evaporation was investigated in the classic work of Hiscock and Weems \cite{HW}, in which it was shown that even if one starts with a rather small value of $Q/M$, the black hole can first approach extremality, before turning around and eventually tends toward the Schwarzschild limit. A different result based on a very different assumption in \cite{wenyu}, on the other hand, showed that these black holes always tend toward the extremal limit.  The point is, in both models, it is possible for a charged black hole to come quite close to its extremal value. It therefore seems that the maximal volumes inside these black holes may actually \emph{decrease}, at least for some time. How then, does one reconcile this with our previous result that the volume is always an increasing function?  

The answer is this: the observation that the volume decreases as the black hole approaches extremality is obtained by
first \emph{fixing} the black hole parameters [say, the mass $M$ and the rotation parameter $a$ in the asymptotically flat Kerr case \cite{1502.01907}], 
then the volume would be [asymptotically] proportional to $v$. The prefactor in front of $v$ then decreases as extremality is approached. However, the point is that, in the dynamical situation, we should not fix the black hole parameters \emph{a priori}, but instead let them evolve over time [under a specific model].

For completeness, let us perform the calculation for an asymptotically flat Reissner-Nordstr\"om black hole with metric 
\begin{equation}
ds^2 = -\left(1-\frac{2M}{r}+\frac{Q^2}{r^2}\right)dt^2 + \left(1-\frac{2M}{r}+\frac{Q^2}{r^2}\right)^{-1} dr^2 + r^2 (d\theta^2 + \sin^2 \theta d\phi^2),
\end{equation}
where $Q>0$ for simplicity. 

The maximal volume inside the black hole is
\begin{flalign}
V &\sim \int^v \int_{\Omega}  \max_r \left[r^2\sin\theta \sqrt{\frac{2M}{r}-\frac{Q^2}{r^2}-1}\right] ~d\theta d\phi dv \\
&= 4\pi \left[\int^v \frac{\sqrt{2}(\sqrt{9M^2-8Q^2}+3M)^2(M\sqrt{9M^2-8Q^2}-3M^2+2Q^2)^{\frac{1}{2}}}{32 Q} ~dv\right],
\end{flalign}
where $M$ and $Q$ are both functions of time.
In order to get to the second line, we have maximized over $r$, which corresponds to choosing the hypersurfaces
\begin{equation}
r^{\pm}=\frac{3M \pm \sqrt{9M^2-8Q^2}}{4}.  
\end{equation}
We choose the positive root since $r^{-} < r_-$, where $r_-$ denotes the inner horizon. [As mentioned in \cite{1411.2854}, one can still consider the region below the inner horizon, but $\partial_r$ is spacelike there, so the contribution to the volume is ``small''.]

That is, the derivative goes like 
\begin{flalign}
\frac{dV}{dv} \sim 4\pi \left[\frac{\sqrt{2}(\sqrt{9M^2-8Q^2}+3M)^2(M\sqrt{9M^2-8Q^2}-3M^2+2Q^2)^{\frac{1}{2}}}{32 Q}\right].
\end{flalign}
 This is \emph{always non-negative}, and tends to zero only when $M \to Q$.

Although an \emph{exactly} extremal black hole would have no ``large volume'', as we have seen, non-extremal black holes continue to have large volumes that grow, no matter how close to extremality they may get. This is not unreasonable since the extremal black holes are very different from their non-extremal cousins\footnote{For example, extremal black holes suffer from instability even at the classical level, but non-extremal ones do not, however close to extremality they are \cite{1206.6598,1307.6800,1212.0729}.}, and furthermore taking the extremal limit is a subtle process \cite{carroll, ingemar, stotyn}.

On the other hand, if we \emph{fix} $M$ and $Q$ to be constant, as in classical general relativity without Hawking evaporation, then the volume grows like 
\begin{equation}
V \sim 4\pi \left[\frac{\sqrt{2}(\sqrt{9M^2-8Q^2}+3M)^2(M\sqrt{9M^2-8Q^2}-3M^2+2Q^2)^{\frac{1}{2}}}{32 Q}\right] v,
\end{equation}
[one may check that in the limit $Q \to 0$, this reduces to Schwarzschild case $V \sim 3\sqrt{3} \pi M^2 v$]
and the prefactor of $v$ decreases monotonically as $M \to Q$.

\addtocounter{section}{4}
\section* {\large{\textsf{4. Conclusion: Maximal Volumes Persist Under Hawking Evaporation}}}

In this work, we have shown with some explicit examples that the maximal volumes inside black holes, first proposed by Christodoulou and Rovelli \cite{1411.2854},  survive the Hawking evaporation process. We also clarified the charged case, in which naively one might suspect that the large interior volume will tend to zero as the black hole evolves toward extremal limit. This is not the case if one properly takes time evolutions of the black hole parameters into account. 

Therefore, we have shown that generically black holes have large interior volumes and that such volumes continue to grow even when the horizon areas shrink under Hawking radiation, confirming the estimate in \cite{1411.2854} that there is sufficient room inside a black hole to store information. This result suggests that black hole remnants may resolve the information loss paradox. Another possibility is that, the black hole completely evaporates and the large interior pinches off as a baby universe. These two possibilities are consistent with the results in \cite{9405007}. 
Of course, either way, as emphasized in \cite{1412.8366}, in order to understand what happens to the information, we have to address what happens to the singularity, this is because any infalling object reaches the singularity in proper time $\tau \leqslant \pi M $ [for an asymptotically Schwarzschild black hole], not withstanding the large interior.

There are nevertheless various loopholes. Firstly, we should not trust the simple ODEs such as Eq.(\ref{ODE1}) to continue to hold near the end stages of Hawking evaporation, as the temperature becomes extremely high \cite{bardeen, hiscock,0412265}. In other words we cannot be confident that something drastic would never happen to the large interior volume. In addition, near the singularity it is conjectured that spacetime is no longer perfectly described by the naive Schwarzschild-interior solution, but rather chaotic, i.e., a BKL singularity \cite{BKL}. [For a recent review, see \cite{1304.6905}].  In such a scenario our analysis would therefore break down. However, for most part of the evolution before these complications arise, our analysis holds --- the maximal volumes inside black holes grow in time, even under Hawking evaporation. Note that, even \emph{if} the volume does continue to grow and thus remains finite when the black hole area shrinks toward zero, such a ``discontinuity'' is not unprecedented --- already in the case of AdS black hole with lens space horizon $S^3/\Bbb{Z}_p$ \emph{at the classical level}, we see that \cite{oyc} the black hole volume is non-vanishing even though the area does, as we take $p \to \infty$. [This, however, is not a dynamical evolution, but rather corresponds to changing the topology of the horizon.] 

Secondly,  is the observation that, if information can indeed be stored in the interior volume of a black hole despite its ever shrinking area, we would have to subscribe to the ``weak form'' interpretation of the Bekenstein-Hawking entropy [that is, contrary to what holography would suggest, the area is \emph{not} the total measure of the information content of the black hole \cite{1412.8366,0901.3156,0003056,0501103}, and so the various entropy bounds \cite{9310026,suss,bekenstein,bousso,9912055} may also be violated]. 

If the volume is in fact the true measure of the information content, the Page time of a black hole would be proportional to its volume instead of its horizon area. Nevertheless, unless the volume is infinite [or, if the volume is finite but Hawking radiation does not carry any information], Page time would eventually set in and one has to deal with potential problems such as the firewall [see sec.(3.3) of \cite{1412.8366}]. Therefore, the suggestion that black hole remnants may have arbitrarily large interior volumes that help ameliorate both the information loss paradox and the firewall paradox is only feasible if Hawking radiation is --- as the original Hawking's calculation suggests --- purely thermal and does not carry any information. For an opposite point of view, see, e.g., \cite{1503.01487}. 

Since other large volume scenarios, such as bubble universes, are most likely not generic [that is, we should not expect these to be inside every black hole \cite{1412.8366}], 
it certainly comes as a relief that general relativity and the simplest model of Hawking radiation with thermal spectrum already provide a generic black hole with large volume that may resolve the information loss paradox.


\addtocounter{section}{1}
\section*{\large{\textsf{Acknowledgement}}}
The author is grateful to Ingemar Bengtsson, Brett McInnes, Michael Good, Sabine Hossenfelder, and Carmen Li for useful comments and discussions. He also thanks Baocheng Zhang for pointing out a crucial mistake in the previous version of this work.


\end{document}